# VM placement over WDM-TDM AWGR PON Based Data Centre Architecture

Azza E. A. Eltraify, Sanaa Hamid Mohamed and Jaafar M.H. Elmirghani

*School of Electronic and Electrical Engineering, University of Leeds, Leeds, LS2 9JT, UK*
*E-mail: {scaeae, elshm, j.m.h.elmirghani}@leeds.ac.uk*

**ABSTRACT**

Passive optical networks (PON) can play a vital role in data centres and access fog solutions by providing scalable, cost and energy efficient architectures. This paper proposes a Mixed Integer Linear Programming (MILP) model to optimize the placement of virtual machines (VMs) over an energy efficient WDM-TDM AWGR PON based data centre architecture. In this optimization, the use of VMs and their requirements affect the optimum number of servers utilized in the data centre when minimizing the power consumption and enabling more efficient utilization of servers is considered. Two power consumption minimization objectives were examined for up to 20 VMs with different computing and networking requirements. The results indicate that considering the minimization of the processing and networking power consumption in the allocation of VMs in the WDM-TDM AWGR PON can reduce the networking power consumption by up to 70% compared to the minimization of the processing power consumption.

## 1. INTRODUCTION

Several studies have focused on the optimization of power efficiency and architectures in data centres and core networks in order to satisfy the high increase in demand for data rates and energy efficiency [1]-[10]. Conventional data centre architecture designs have faced many challenges over the past few decades which resulted in the need to develop new designs that are capable of providing more scalable, reliable and efficient data centre architectures [11]-[18], [22]-[41].

With the growth of data centres and the increasing number of power-hungry devices within them, a need for designs with passive components has risen to provide more energy efficient architectures with better resource utilization and lower cost [19]. Passive optical networks were introduced as a solution to several challenges in data centre and core networks, which led to enhanced performance in access networks while lowering the cost and latency, increasing the capacity, scalability and providing overall energy efficiency. The passive devices used in these architecture designs are Arrayed Waveguide Grating Routers (AWGR), Fibre Bragg gratings (FBG), and star couplers/splitters [20].

One of the main challenges in data centre architecture design is the underutilization of resources due to the use of ineffective resource allocation algorithms such as round robin. To tackle this issue, virtualization can be used alongside resource provisioning algorithms to mitigate the non-energy efficient utilization of resources within the network.

This paper proposes a Mixed Integer Linear Programming (MILP) model to optimize the placement of virtual machines (VMs) over an energy efficient WDM-TDM AWGR PON based data centre architecture. This model efficiently maps VMs to servers to ensure the most effective utilization of servers, which lowers the number of active servers hence reducing the power consumption.

The remainder of this paper is organized as follows, Section 2 discusses the WDM-TDM PON based data centre architecture, Section 3 discusses the optimization model, Section 4 discusses the results while Section 5 concludes the paper with an overall summary.

## 2. WDM-TDM PON-BASED DATA CENTRE ARCHITECTURE

Figure.1 shows the WDM-TDM AWGR PON-based data centre architecture, which is a PON cell consisting of 4 racks holding 7 servers each. Communication among racks is provisioned through two intermediate AWGRs through 4 different wavelengths. Another set of 1:N AWGRs provision the communication with the OLT. Each server is equipped with photodetectors and tuneable lasers to enable optimum wavelength selection. Inter-rack communication is achieved either through the AWGR or the OLT, where the AWGR relay traffic to the designated server according to the selected wavelength. All routes available are kept in a routing interconnection map, which facilities the ability to find alternative routes if required hence providing multi-path routing and load balancing [11].

WDM-TDM-PON was deployed in this architecture to optimize the connections within the architecture whether among the OLT and PON groups or within the PON groups. Introducing TDM to the WDM architecture enabled better utilization of resources where several wavelengths can be sent using different time slots to different destination, providing full connectivity among the OLT, ONU and PON groups [21].

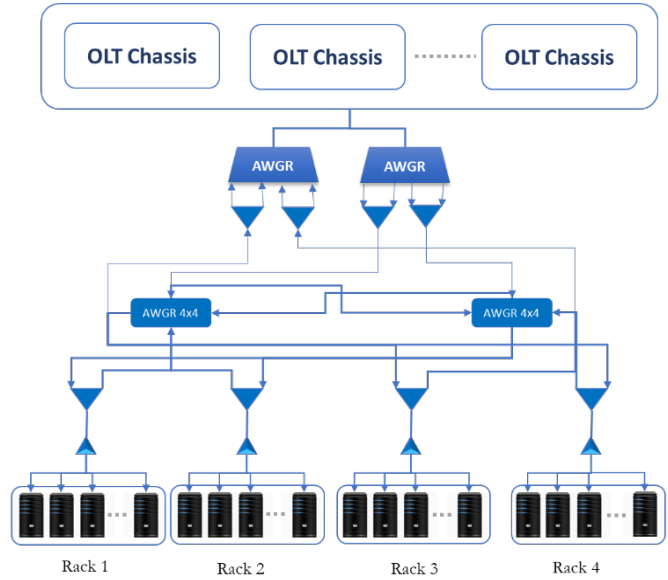

*Figure 1. AWGR-Based Passive Optical Network Data Centre Architecture*

## 3. The OPTIMIZATION MODEL

A linear mathematical programming approach was used in this model to mitigate the effects of the underutilized resources by optimally provisioning resources through a resource allocation algorithm. Several conventional algorithms such as the Best Fit Deceasing Bin-Packing have tackled this issue. Our approach is to minimize the power consumption by the use of a MILP model to optimize the virtualization, where VMs are mapped into servers and only utilized when requested.

The objective of this MILP optimization model is to minimize the power consumption of the WDM-TDM AWGR PON-based data centre architecture. Several parameters are considered in this model such as the number of VMs, inter VM traffic, and the processing requirements of the VMs. In addition, consideration is given to the processing capacity of each server and the capacity of each server's ONU. The traffic and demands between the VMs is randomly generated using a uniform distribution between 0.1Gbps and 4 Gbps. The processing capacities of the total 28 servers are also randomly generated between 1.8 GHz and 2.75 GHz with uniform distribution.

Moreover, a constraint is applied to limit the number of servers mapped to VMs at a certain time. Another set of constraints are applied to facilitate the communication among the PON groups and within them, to ensure that the traffic among different groups is within the capacity of the utilized wavelength. The CPLEX solver was used to provide the solution and the model file is written using AMPL. Table 1 below summarizes the key MILP model parameters.

*Table 1. Key parameters for the MILP model.*

| Parameter | Value |
| --- | --- |
| Maximum power consumption of a server [14] | 301 Watt |
| Idle power consumption of a server [14] | 201 Watt |
| Processing capacity of a server | 1.8 GHz - 2.75 GHz |
| Number of servers per rack | 7 |
| Number of racks | 4 |
| Power consumption of an ONU [14] | 2.5 Watt |
| Maximum upload and download capacities of the ONU | 10 Gbps |
| Number of VMs | 5 VMs - 20 VMs |
| Processing requirements of a VM | 0.1 GHz - 0.5 GHz |
| Inter VMs traffic | 0.1Gbps - 4 Gbps |

## 4. RESULTS

In this paper, the WDM-TDM AWGR PON-based data centre architecture was modelled to provision the resources and optimize the VMs allocation with the objective of minimizing the power consumption. The model was run using different numbers of VM requests, and different VMs processing requirements. The number of VM requests used were 5, 10,15, and 20 requests.

The results showed that the model favours allocating several VMs to the same server as long as it meets the capacity constraints, which reduces the levels of traffic flow between servers in the network and the amount of power consumed. This results in a number of servers being switched off or idle, hence lowering the levels of power consumption.

The model was run with two objectives. The first considers minimizing the processing power consumption (Pc) only while the second considers minimizing the processing and networking power consumption (Pc + Pn). Figure 2 (a) and (b) show the total power consumption and the networking power consumption, respectively, when considering the two objectives for 5, 10, 15, and 20 VMs. In this comparison, the average processing requirement for each VM is 0.3 GHz. For such number of VMs and relatively low processing requirements, most of the allocations were fit in a server or two which leads to low networking power consumption and hence both objectives achieve similar total power consumption. Figure 3 (a) and (b) show the total power consumption and the total networking power consumption results for the case of 10 VMs while increasing processing requirement per VM. As this requirement increases, more servers are needed to serve each VM which leads to utilizing more servers and having increased inter VMs traffic between the servers. The total power consumption results under the two objectives show that the allocation is different for each, hence, up to 70% difference in the networking power consumption.

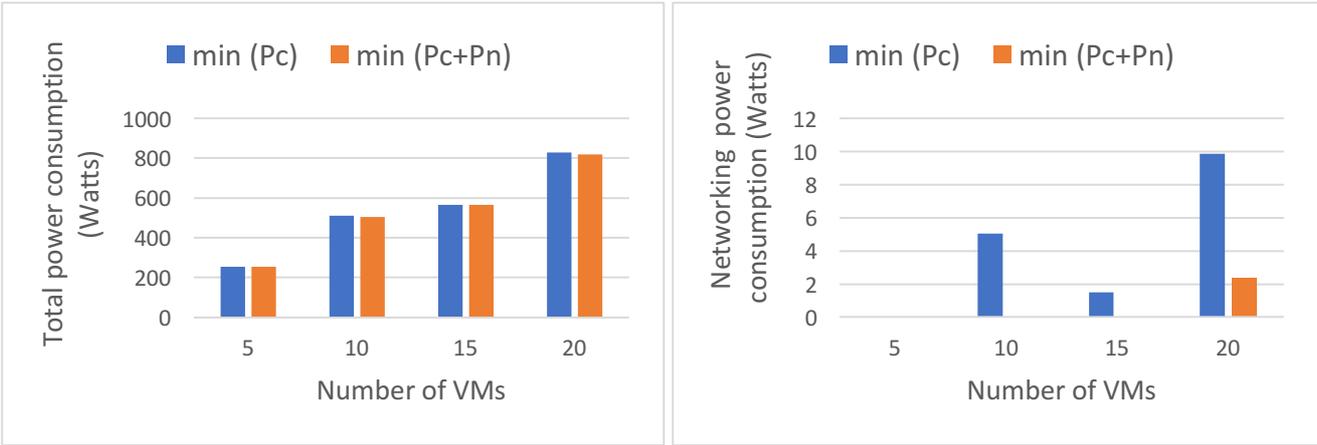

*Figure 2. (a) Total power Consumption and (b) Networking power consumption, for different VM numbers*

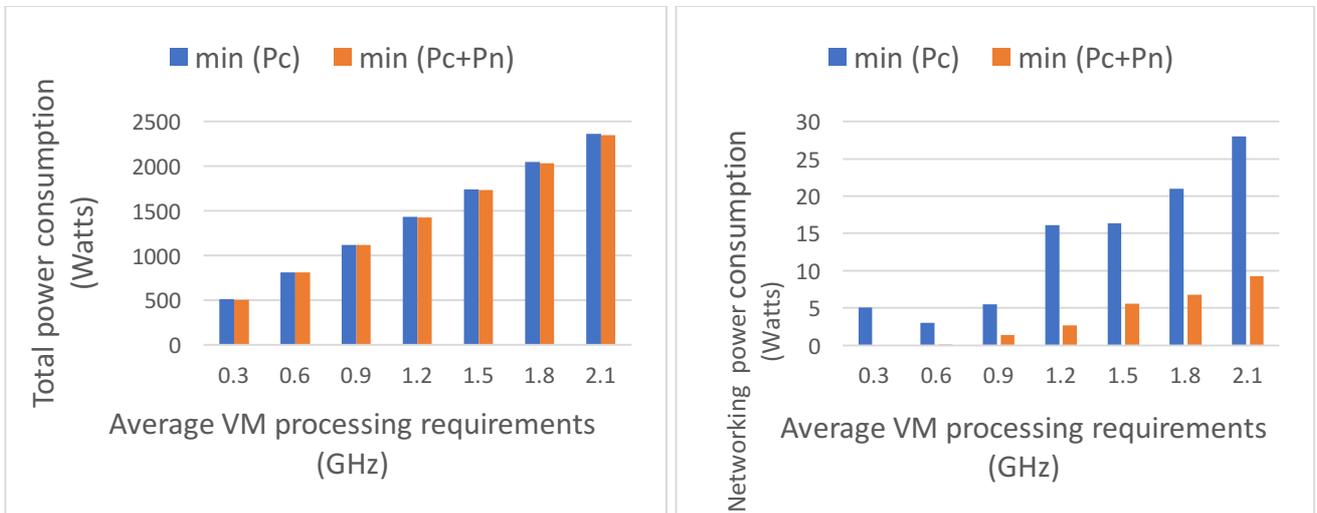

*Figure 3. (a) Total, and (b) Networking power consumption of 10 VMs with different average processing requirements.*

## 5. CONCLUSIONS

This paper introduced a mathematical optimization model to optimize the mapping of virtual machines to servers in a WDM-TDM AWGR PON-based data centre architecture. Several VMs were mapped to a single server to ensure that not all the servers in the network are utilized at the same time, which resulted in lower power consumption. The effects of having different numbers of virtual machines and an increasing processing requirement on power consumption has also been tested. Two energy efficiency objectives were also examined and compared. The results show that minimizing the processing and networking power consumption can reduce the networking power consumption by up to 70%. Future work includes considering the memory requirements of the VMs and its impact on the placements and to consider higher number of VMs.

ACKNOWLEDGEMENTS

The authors would like to acknowledge funding from the Engineering and Physical Sciences Research Council (EPSRC), INTERNET (EP/H040536/1), STAR (EP/K016873/1) and TOWS (EP/S016570/1) project. All data are provided in full in the results section of this paper.

**REFERENCES**

[1] X. Dong, T. El-Gorashi, and J. M. Elmirghani, "IP over WDM networks employing renewable energy sources," *J. Lightwave Technol.,* vol. 29, no. 1, pp. 3–14, 2011.

[2] X. Dong, T. El-Gorashi, and J. M. H. Elmirghani, "Green IP over WDM networks with data centers," *J. Lightwave Technol.,* vol. 29, no. 12, pp. 1861–1880, June 2011.

[3] X. Dong, T. E. H. El-Gorashi, and J. M. H. Elmirghani, "On the energy efficiency of physical topology design for IP over WDM networks," *J. Lightwave Technol.,* vol. 30, no. 12, pp. 1931–1942, 2012.

[4] A. Lawey, T. El-Gorashi, and J. Elmirghani, "Distributed energy efficient clouds over core networks," *J. Lightwave Technol.,* vol. 32, no. 7, pp. 1261–1281, Apr. 2014.

[5] N. Osman, T. El-Gorashi, L. Krug, and J. Elmirghani, "Energy-efficient future high-definition TV," *J. Lightwave Technol.,* vol. 32, no. 13, pp. 2364–2381, July 2014.

[6] A. Lawey, T. El-Gorashi, and J. Elmirghani, "BitTorrent content distribution in optical networks," *J. Lightwave Technol.,* vol. 32, no. 21, pp. 4209–4225, Nov. 2014.

[7] L. Nonde, T. El-Gorashi, and J. Elmirghani, "Energy efficient virtual network embedding for cloud networks," *J. Lightwave Technol.,* vol. 33, no. 9, pp. 1828–1849, May 2015.

[8] M. Musa, T. Elgorashi, and J. Elmirghani, "Energy efficient survivable IP-over-WDM networks with network coding," *J. Opt. Commun. Netw.* vol. 9, no. 3, pp. 207–217, 2017.

[9] Musa, Mohamed, Taisir Elgorashi, and Jaafar Elmirghani. "Bounds for Energy-Efficient Survivable IP Over WDM Networks With Network Coding" *Journal of Optical Communications and Networking* 10.5 (2018): 471-481, 2018.

[10] Elmirghani, J. M. H., T. Klein, K. Hinton, L. Nonde, A. Q. Lawey, T. E. H. El-Gorashi, M. O. I. Musa, and X. Dong. "GreenTouch GreenMeter core network energy-efficiency improvement measures and optimization" *Journal of Optical Communications and Networking* 10, no. 2 (2018): A250-A269.

[11] Ali Abdullah Hammadi: "Future PON Data Centre Networks", *University of Leeds, School of Electronic and Electrical Engineering*, Aug. 2016.

[12] J. Beals IV, N. Bamiedakis, A. Wonfor, R. Penty, I. White, J. DeGroot Jr, et al.: "A terabit capacity passive polymer optical backplane based on a novel meshed waveguide architecture", *Applied Physics A,* vol. 95, pp. 983-988, 2009.

[13] Hammadi, Ali, Taisir EH El-Gorashi, Mohamed OI Musa, and Jaafar MH Elmirghani, "Server-centric PON data center architecture" *In Transparent Optical Networks (ICTON), 2016 18th International Conference on,* pp. 1-4. IEEE, 2016.

[14] A. Hammadi, Mohamed O. I. Musa, T. E. H. El-Gorashi, and J.M.H. Elmirghani, "Resource Provisioning for Cloud PON AWGR-Based Data Center Architecture", *21st European Conference on Network and Optical Communications (NOC),* Portugal, 2016.

[15] Hammadi, Ali, Taisir EH El-Gorashi, and Jaafar MH Elmirghani. "High performance AWGR PONs in data centre networks", *In Transparent Optical Networks (ICTON), 2015 17th International Conference on,* pp. 1-5. IEEE, 2015.

[16] John W. Lockwood, Nick McKeown, Greg Watson, Glen Gibb, Paul Hartke, Jad Naous, Ramanan Raghuraman, and Jianying Luo; "NetFPGA - An Open Platform for Gigabit-rate Network Switching and Routing"; *MSE 2007,* San Diego, June 2007.

[17] A. Hammadi, T. E. El-Gorashi, and J. M. H. Elmirghani; "PONs in Future Cloud Data Centers"; *IEEE Communications Magazine.*

[18] D. Kliazovich, P. Bouvry, and S. U. Khan, "GreenCloud: a packet-level simulator of energy-aware cloud computing data centers," *The Journal of Supercomputing,* vol. 62, pp. 1263-1283, 2012.

[19] C. Kachris and I. Tomkos, "Power consumption evaluation of hybrid WDM PON networks for data centers," in Networks and Optical Communications (NOC), 2011 *16th European Conference on*, 2011, pp. 118-121.

[20] Kani, Jun-ichi. "Enabling technologies for future scalable and flexible WDM-PON and WDM/TDM-PON systems." *IEEE Journal of Selected Topics in Quantum Electronics 16,* no. 5 (2010): 1290-1297.

[21] Eltraify, Azza EA, Mohamed OI Musa, Ahmed Al-Quzweeni, and Jaafar MH Elmirghani. "Experimental Evaluation of Passive Optical Network Based Data Centre Architecture." In *2018 20th International Conference on Transparent Optical Networks (ICTON)*, pp. 1-4. IEEE, 2018.

[22] M. Musa, T.E.H. El-Gorashi and J.M.H. Elmirghani, "Bounds on GreenTouch GreenMeter Network Energy Efficiency," *IEEE/OSA Journal of Lightwave Technology*, vol. 36, No. 23, pp. 5395-5405, 2018.

[23] H.M.M., Ali, A.Q. Lawey, T.E.H. El-Gorashi, and J.M.H. Elmirghani, "*Future Energy Efficient Data Centers With Disaggregated Servers*," IEEE/OSA Journal of Lightwave Technology, vol. 35, No. 24, pp. 5361 – 5380, 2017.

[24] B. Bathula, M. Alresheedi, and J.M.H. Elmirghani, "*Energy efficient architectures for optical networks*," Proc IEEE London Communications Symposium, London, Sept. 2009.

[25] B. Bathula, and J.M.H. Elmirghani, "*Energy Efficient Optical Burst Switched (OBS) Networks*," IEEE GLOBECOM'09, Honolulu, Hawaii, USA, November 30-December 04, 2009.

[26] X. Dong, T.E.H. El-Gorashi and J.M.H. Elmirghani, "*Green Optical OFDM Networks,*" IET Optoelectronics, vol. 8, No. 3, pp. 137 – 148, 2014.
[27] M. Musa, T.E.H. El-Gorashi and J.M.H. Elmirghani, "*Energy Efficient Survivable IP-Over-WDM Networks With Network Coding,*" IEEE/OSA Journal of Optical Communications and Networking, vol. 9, No. 3, pp. 207-217, 2017.
[28] A.M. Al-Salim, A. Lawey, T.E.H. El-Gorashi, and J.M.H. Elmirghani, "*Energy Efficient Big Data Networks: Impact of Volume and Variety,*" IEEE Transactions on Network and Service Management, vol. 15, No. 1, pp. 458 - 474, 2018.
[29] A.M. Al-Salim, A. Lawey, T.E.H. El-Gorashi, and J.M.H. Elmirghani, "Greening big data networks: velocity impact," *IET Optoelectronics,* vol. 12, No. 3, pp. 126-135, 2018.
[30] N. I. Osman and T. El-Gorashi and J. M. H. Elmirghani, "The impact of content popularity distribution on energy efficient caching," in *2013 15th International Conference on Transparent Optical Networks (ICTON)*, June 2013, pp. 1-6.
[31] L. Nonde, T. E. H. El-Gorashi, and J. M. H. Elmirghani, "Virtual Network Embedding Employing Renewable Energy Sources," *in 2016 IEEE Global Communications Conference (GLOBECOM)*, Dec 2016, pp. 1-6.
[32] A. Q. Lawey, T. E. H. El-Gorashi, and J. M. H. Elmirghani, "Renewable energy in distributed energy efficient content delivery clouds," in *2015 IEEE International Conference on Communications (ICC),* June 2015, pp. 128-134.
[33] A.N. Al-Quzweeni, A. Lawey, T.E.H. El-Gorashi, and J.M.H. Elmirghani, "Optimized Energy Aware 5G Network Function Virtualization," *IEEE Access*, vol. 7, pp. 44939 - 44958, 2019.
[34] M.S. Hadi, A. Lawey, T.E.H. El-Gorashi, and J.M.H. Elmirghani, "Patient-Centric Cellular Networks Optimization using Big Data Analytics," *IEEE Access*, vol. 7, pp. 49279 - 49296, 2019.
[35] M.S. Hadi, A. Lawey, T.E.H. El-Gorashi, and J.M.H. Elmirghani, "Big Data Analytics for Wireless and Wired Network Design: A Survey, *Elsevier Computer Networks,* vol. 132, No. 2, pp. 180-199, 2018.
[36] S. Igder, S. Bhattacharya, and J. M. H. Elmirghani, "Energy Efficient Fog Servers for Internet of Things Information Piece Delivery (IoTIPD) in a Smart City Vehicular Environment," in *2016 10th International Conference on Next Generation Mobile Applications, Security and Technologies (NGMAST),* Aug 2016, pp. 99-104.
[37] S. H. Mohamed, T. E. H. El-Gorashi, and J. M. H. Elmirghani, "Energy Efficiency of Server-Centric PON Data Center Architecture for Fog Computing," in *2018 20th International Conference on Transparent Optical Networks (ICTON),* July *2018*, pp. 1-4.
[38] A. E. A. Eltraify, M. O. I. Musa and J. M. H. Elmirghani, "TDM/WDM over AWGR Based Passive Optical Network Data Centre Architecture," *2019 21st International Conference on Transparent Optical Networks (ICTON)*, Angers, France, 2019, pp. 1-5.
[39] A. E. A. Eltraify, M. O. I. Musa, A. Al-Quzweeni and J. M. H. Elmirghani, "Experimental Evaluation of Server Centric Passive Optical Network Based Data Centre Architecture," *2019 21st International Conference on Transparent Optical Networks (ICTON)*, Angers, France, 2019, pp. 1-5.
[40] X. Dong, A.Q. Lawey, T.E.H. El-Gorashi, and J.M.H. Elmirghani, "Energy Efficient Core Networks," *Proc 16[th] IEEE Conference on Optical Network Design and Modelling* (*ONDM'12*), 17-20 April, 2012, UK.
[41] Al-Azez, Z., Lawey, A., El-Gorashi, T.E.H., and Elmirghani, J.M.H., "Energy Efficient IoT Virtualization Framework with Peer to Peer Networking and Processing" *IEEE Access*, vol. 7, pp. 50697 - 50709, 2019.